\documentclass[journal=jaccck,manuscript=article, layout=twocolumn]{achemso}
\usepackage[version=3]{mhchem} % Formula subscripts using \ce{}
\usepackage[T1]{fontenc}       % Use modern font encodings

\usepackage[version=3]{mhchem} % Formula subscripts using \ce{}
\usepackage[T1]{fontenc}       % Use modern font encodings

\usepackage{array}
\usepackage{booktabs}

\usepackage{listings}
\usepackage{lmodern}
\usepackage{caption}
\usepackage{float}
\usepackage{geometry}
\usepackage{setspace}
\usepackage{hyperref}
\usepackage{amsmath}
\usepackage{bm}
\usepackage[lofdepth,lotdepth]{subfig}
\usepackage{color}
\newcommand{\im}{\mathrm{Im}}
\author{K. Beltako}
\author{F. Michelini}
\email{fabienne.michelini@im2np.fr}
\author{N. Cavassilas, L. Raymond}
\affiliation{Aix-Marseille University, CNRS, IM2NP, UMR 7334, 13288 Marseille, France}
\title{Dynamical photo-induced electronic properties of molecular junctions}

\begin{document}

\begin{tocentry}
\begin{center}
 \includegraphics[width=8.5 cm]{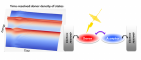}
\end{center}
\end{tocentry}

%%%%%%%%%%%%%%%%%%%%%%%%%%%%%%%%%%%%%%%%%%%%%%%%%%%%%%%%%%%%%%%%%%%%%
% The "tocentry" environment can be used to create an entry for the
% graphical table of contents. It is given here as some journals
% require that it is printed as part of the abstract page. It will
% be automatically moved as appropriate.
%%%%%%%%%%%%%%%%%%%%%%%%%%%%%%%%%%%%%%%%%%%%%%%%%%%%%%%%%%%%%%%%%%%%
%\begin{tocentry}
%
%Some journals require a graphical entry for the Table of Contents.
%This should be laid out ``print ready'' so that the sizing of the
%text is correct.
%
%Inside the \texttt{tocentry} environment, the font used is Helvetica
%8\,pt, as required by \emph{Journal of the American Chemical
%Society}.
%
%The surrounding frame is 9\,cm by 3.5\,cm, which is the maximum
%permitted for  \emph{Journal of the American Chemical Society}
%graphical table of content entries. The box will not resize if the
%content is too big: instead it will overflow the edge of the box.
%
%This box and the associated title will always be printed on a
%separate page at the end of the document.
%
%
%\end{tocentry}

\begin{abstract}
Nanoscale molecular-electronic devices and machines are emerging as promising functional elements, naturally flexible and efficient, for next-generation technologies. A deeper understanding of carrier dynamics in molecular junctions is expected to benefit many fields of nanoelectronics and power-devices.
We determine time-resolved charge current flowing at donor-acceptor interface in molecular junctions connected to metallic electrodes by means of quantum transport simulations. The current is induced by the interaction of the donor with a Gaussian-shape femtosecond laser pulse.  Effects of the molecular internal coupling, metal-molecule tunneling and light-donor coupling on photocurrent are discussed. 
We then examine the junction working through the time-resolved donor density of states. Non-equilibrium reorganization of hybridized molecular orbitals through the light-donor interaction gives rise to two phenomena: the dynamical Rabi shift and the appearance of Floquet-like states. 
Such insights into the dynamical photoelectronic structure of molecules are of strong interest for ultrafast spectroscopy, and open avenues toward the possibility of analyzing and controlling the internal properties of quantum nanodevices with pump-push photocurrent spectroscopy.
\end{abstract}

%%%%%%%%%%%%%%%%%%%%%%%%%%%%%%%%%%%%%%%%%
%%%%%%%%%%%%%%%%%%%%%%%%%%%%%%%%%%%%%%%%%
\section{Introduction}
%%%%%%%%%%%%%%%%%%%%%%%%%%%%%%%%%%%%%%%%%
%%%%%%%%%%%%%%%%%%%%%%%%%%%%%%%%%%%%%%%%%

Recent striking theoretical and experimental advances in electronic transport through molecular wires and junctions have generated considerable interest for high frequency quantum transport. ~\cite{mentovich_gated-controlled_2013, diez-perez_ambipolar_2012,xu_large_2005,lortscher_transport_2012}
Molecular electronic devices form a  promising alternative to standard switches due to their ultrashort
response time, low cost, efficiency and flexible nature.
Time-dependent numerical investigations may offer a high degree of flexibility to promote the realization of such systems. 
%Time-dependent numerical investigations may promote the realization of such systems. 
Arbitrary temporal shapes of interacting potentials can be easily incorporate into dynamical simulations,~\cite{selzer_transient_2013, volkovich_transient_2011} allowing for the description of alternating and transient currents, and, hence, the optimization of switching processes.
One can also deal with time-dependent external fields that modulate the transport characteristics, and thus investigate photo-induced and photo-assisted transport,~ \cite{platero_photon-assisted_2004} as well as molecular photo-switches. 
Time-dependent investigations also afford the possibility to develop highly resolved ultrafast scanning probe.~\citep{cocker_ultrafast_2013, pivrikas_review_2007, brauer_dynamics_2015, bakulin_charge_2013}
The idea of measuring the internal ultrafast dynamics of molecular junctions through photocurrent, instead of emission,~\citep{bakulin_ultrafast_2016}  has shown to reveal underlying physics of molecular junctions, as the key process of charge transfert and separation at interfaces~\citep{jakowetz_what_2016, ono_minimal_2016} also involved in energy conversion. Indeed, time-dependent investigations would also bring new perspectives for  high conversion efficiency for solar cells with molecular blends, as suggested by numerical simulations using quantum dots or bulk heterojunction as models~\citep{prins_fast_2012,alsulami_remarkably_2016} or non-equilibrium stationary transport studies.~\citep{cavassilas_modeling_2013,beltako_state_2016}
%
%The use of  quantum dots  and bulk heterojunction~\citep{prins_fast_2012,alsulami_remarkably_2016} as model for these numerical simulations have also suggested high conversion efficiency for solar cells with molecular blends~\citep{alsulami_remarkably_2016} in time-resolved cases compared to the non-equilibrium stationary transport.~\citep{cavassilas_modeling_2013,beltako_state_2016}
Further issues to be handled in this direction will be energy transfer and entropy,~\citep{michelini_entropy_2017} molecular vibrations, and electron-hole interactions.~\citep{nemati_aram_modeling_2016,nemati_aram_impact_2017}
However, limitations and challenges still exist for the development of molecular electronics. Characterization techniques and
theoretical simulations should be valuable for deeply understanding charge transport through molecular junctions, and, hence, for conducting relevant device design and future directions of molecular electronics.~\cite{xiang_molecular-scale_2016,evers_preface:_2017}
In particular, correlation of the photocurrent directionality~\citep{kuperman_field-induced_2017} to the field shape and the internal molecular structure has to be deepened. It is here crucial to decipher the intimate nature of such quantum systems, which is provided by features of their dynamical field-induced electronic structure, like the occurence of sidebands~\citep{hertel_ultrafast_2006,compton_dynamic_2011} or energy level shift.~\citep{cochrane_pronounced_2015}

In this paper, we investigate the dynamical photoelectronic properties induced by a Gaussian-shape femtosecond laser pulse in a donor-acceptor junction laterally connected to two metallic electrodes. We rely on quantum  transport simulations in the framework of the Keldysh's formalism, using a few-level quantum model of the donor-acceptor molecule. Numerical analysis makes possible to correlate the time- and energy-resolved intramolecular dynamics to the photocurrent flowing at the D/A interface.

 %%%%%%%%%%%%%%%%%%%%%%%%%%%%%%%%%%%%%%%%%
%%%%%%%%%%%%%%%%%%%%%%%%%%%%%%%%%%%%%%%%%
\section{Theoretical methods} 
 %%%%%%%%%%%%%%%%%%%%%%%%%%%%%%%%%%%%%%%%%
%%%%%%%%%%%%%%%%%%%%%%%%%%%%%%%%%%%%%%%%%

Among the standard remarkable methods of quantum statistics, non-equilibrium Green's function (NEGF) formalism has already shown its ability to simulate different types of nanodevices~ \cite{lake_nonequilibrium_1992}. Derived from this formalism, an alternative approach, called wave function (WF) technic, was recently developed~\citep{gaury_numerical_2014} and  seems adequate for the numerical simulation of time-resolved transport in quantum systems.
As NEGFs, WF technic also originates from the theoretical framework developed by Keldysh to tackle many-body problems~\cite{settnes_patched_2015}. In WF technic, building blocks are wave functions (vectors) instead of Green's functions (matrices). One of these wave functions represents the projection of the Green's function on one of the incoming modes propagating from reservoirs to the system. Switching from matrices to vectors relaxes the computational burden of time-dependent problems. Nonetheless, WF technic for scattering mechanisms beyond mean-field approximation still remains undeveloped.

Let us consider an open quantum system $S$, connected to two semi-infinite leads at left and right sides, $L$ and $R$. The total Hamiltonian is written
\begin{eqnarray}
\label{eqHamil}
H^{tot}(t)&=& \sum_{i,j\in S} H_{ij}(t)c_{i}^{\dagger}c_{j}+\sum_{i,j\in L,R} R_{ij}c_{i}^{\dagger}c_{j}\nonumber\\
&&+\sum_{i\in S,j\in L,R} T_{ij}c_{i}^{\dagger}c_{j}+h.c. \, ,
\end{eqnarray}
where $c_{i}^{\dagger}(c_{i})$ is the creation (annihilation) operator for a single particle on site \emph{i}. Elements $H_{ij}$ and $R_{ij} $  stand respectively for system ($S$) and leads ($L,R$). Coupling $T_{ij}$  is the system-lead tunneling parameter from site $i$ in $S$ to site $j$ in $L$ or $R$. In the following, we will used matrix notation in bold style. Matrix elements of the system $H_{ij}$ form $\bf{H}(t)$, which is assumed finite. In NEGF formalism, scattering to semi-infinite leads is encoded in self-energies ${\bf\Sigma}_{L,R}$.

From NEGF formalism to WF technic, the mathematical recipe is the diagonalisation of these time-independent self-energies:
\begin{eqnarray}
\label{selfE}
{\bf\Gamma}_{L,R}(E)&=&\sum_{\alpha}v_{\alpha}(E){\boldsymbol \xi}_{\alpha E}{\boldsymbol \xi}_{\alpha E}^{\dagger}\, ,
\end{eqnarray}
where ${\boldsymbol \xi}_{\alpha E}$ are the transverse modes at energy $E$ and velocity $v_{\alpha}$  ($\alpha$ account for the different modes from all leads). In a nutshell, the expression of the lesser self-energy,
 \begin{eqnarray}
 \label{selft}
 {\bf\Sigma}^{<}(t-t')=\sum_{\alpha\in L,R}\int \frac{dE}{2\pi}if_{\alpha}(E)e^{-i\frac{E}{\hbar}(t-t')}{\bf \Gamma}_{\alpha}(E)\, ,
 \end{eqnarray}
is inserted into the integrated equation of motion of the lesser Green's function,
 \begin{eqnarray}
 {\bf G}^{<}(t,t')=\int du dv  {\bf G}^{R}(t,u){\bf \Sigma}^{<}(u,v)[{\bf G}^{R}(v,t')]^{\dagger} \, .
 \end{eqnarray}
It follows an expression of $G^{<}$ in terms of WFs:
\begin{eqnarray}
G^{<}_{ij}(t,t')=\sum_{\alpha}\int \frac{dE}{2\pi}f_{\alpha}(E){\Psi}_{\alpha E}(i,t) {\Psi}_{\alpha E}^{\dagger}(j,t') \, ,
\end{eqnarray}
with $f_{\alpha}(E)$ the Fermi-Dirac distribution function, and ${\bf \Psi}_{\alpha E}(i,t)$ the $i^{th}$ component of the WF defined as
\begin{eqnarray}
{\bf \Psi}_{\alpha E}(t)=\sqrt{v_{\alpha}}\int du e^{-iEu/\hbar} {\bf G}^{r}(t,u){\boldsymbol \xi}_{\alpha E}\, .
\end{eqnarray}
 
 A widely used strategy to include time-dependence is to separate the hamiltonian into a stationary known problem  $\bf{H}_{0}$ and a time-dependent
perturbation ${\bf H}_{p}(t)$: ${\bf H}(t) = {\bf H}_{0} + {\bf H}_{p}(t)$.
 The stationary transport equation is the usual bare Green's function equation of motion projected onto the lead modes, which is given below as $\alpha$ linear sparse equations:
  \begin{align}
  \label{PHIST}
  [E{\bf I}-{\bf H}_{0}-{\bf \Sigma}^{r}(E)]{\bf \Psi}_{\alpha E}^{st}=\sqrt{v_{\alpha}(E)}{\boldsymbol \xi}_{\alpha E}\, ,
  \end{align}
where  ${\bf \Psi}_{\alpha E}^{st}$ is the stationary wave function, $\bf{I}$ is the identity matrix and  ${\bf \Sigma}^{r}$ the retarded self-energy of the leads.
When  $ {\bf H}_{p}(t)$ is switched on, the system is consistently described by WFs, ${\bf \Psi}_{\alpha E} (t)$, which also split into
\begin{align}
{\bf \Psi}_{\alpha E} (t)= {\bf \Psi}^{p}_{\alpha E}(t) +e^{-iEt/\hbar}{\bf \Psi}_{\alpha E}^{st} \, .
\end{align}  
 Here ${\bf \Psi}^{p}_{\alpha E}(t)$ is a wave vector that measures the time-dependent deviation from ${\bf \Psi}_{\alpha E}^{st}$.  
 Each ${\bf \Psi}^{p}_{\alpha E}(t)$ complies with the equation of motion of the retarded Green's function by being the solution of the integro-differential equation:
\begin{eqnarray}
\label{PHIP}
  i\hbar\partial_{t} {\bf \Psi}^{p}_{\alpha E}(t)&=&{\bf H}(t) {\bf \Psi}^{p}_{\alpha E}(t) \nonumber\\
  &+&\int_{0}^{t}du {\bf \Sigma}^{r}(t-u) {\bf \Psi}^{p}_{\alpha E}(u) \nonumber\\
  &+& {\bf H}_{p}(t)e^{-iEt/\hbar} {\bf \Psi}_{\alpha E}^{st} \, .
\end{eqnarray}  
  
Moreover, in mesoscopic systems with energy scales smaller that of the Fermi energy variation in electrodes (like in the case of metallic electrodes), the energy dependence of lead self-energy might be neglected to break the non-locality in time and the memory kernel of self-energy. This is known as wide band limit (WBL) approximation, which enables one to write ${\bf \Sigma}^{r}(t-u)=i{\bf \Gamma} \delta(t-u)$, and thus simplifies eq~\ref{PHIP}. Numerical calculations were done in the WBL approximation.
%Each $ \Gamma$ is called tunneling rate and represent the coupling of the electrodes to the connected sites of the system (molecular junction).

% A summary of the fastest numerical simulation algorithm as given in the original paper~\citep{gaury_numerical_2014} are the followings. First we define the total Hamiltonian of the system, construct the stationnary modes, compute the self-energy of the leads. Next solve the two transport equations \ref{PHIST} and \ref{PHIP} to reconstruct the full wave function of the system.
Finally, all physical observables can be computed from the full set of WFs, like the charge current flowing from site $i$ to site $j$:
\begin{eqnarray}
I_{ij}(t)&=&-\dfrac{2e} {\hbar} \im \Big[\sum_{\alpha}\int \frac{dE}{2\pi}f_{\alpha}(E) \nonumber \\
&& \times \Psi_{\alpha E}^{\dagger}(i,t) H_{ij}(t)\Psi_{\alpha E}(j,t) \Big]\, .
\end{eqnarray}  
 
\begin{figure}
   \begin{center}
      \begin{tabular}{c} 
	\includegraphics[width=8.25 cm]{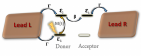}\\
      	\includegraphics[width=5.5 cm]{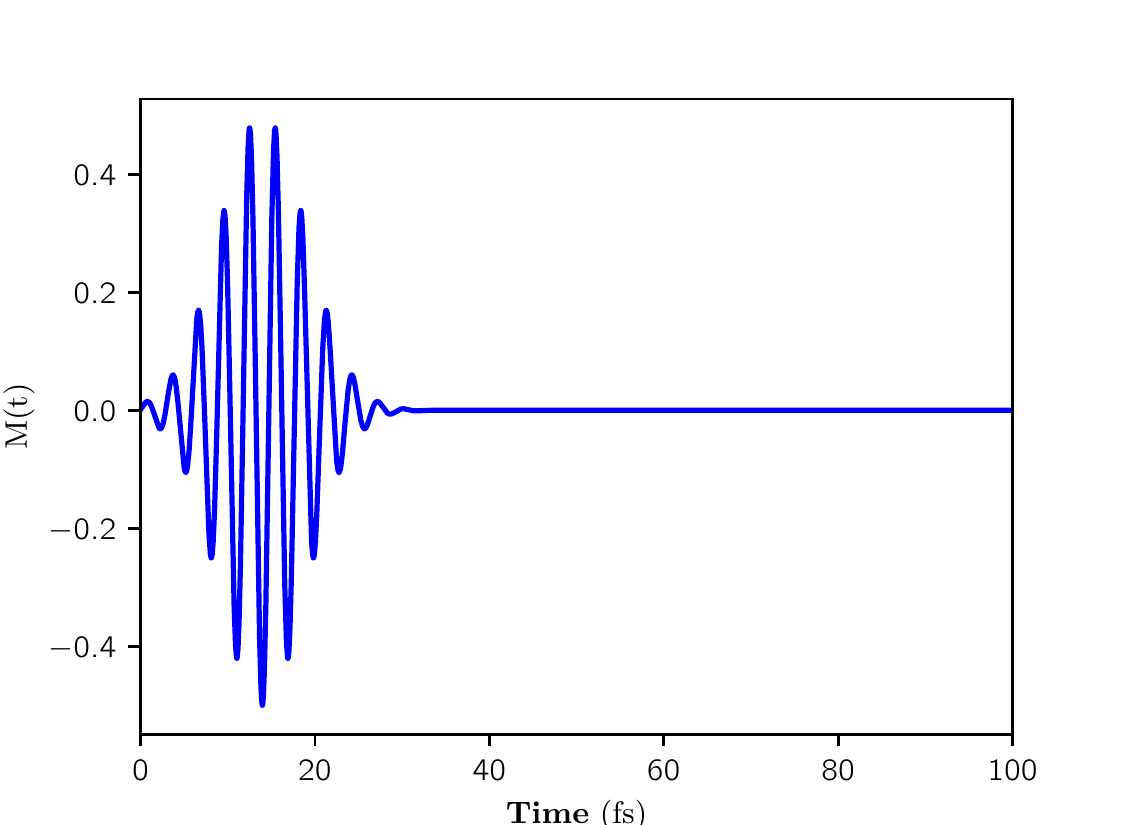}
     \end{tabular}
   \end{center}
   \caption 
   {\label{systa} Schematic representation of the molecular junction (upper panel). Pulse-induced light-donor coupling as a function of time $M(t)$ (lower panel). }
   \end{figure}

 In the present study, we consider an unbiased molecular junction laterally in contact with two metallic electrodes, as schematically depicted in Figure~\ref{systa}.
The molecular complex consists of a donor-bridge-acceptor chain, which represents a key molecular model for biological applications. We simplify the overall device as a three-level system made of a donor D (two levels) connected to an acceptor A (a single level) through a coupling parameter $\beta$ that globally describes the bridge role. The molecular complex ($S$) is connected on the left and the right to metallic leads.
In this model, ${\bf H}_{0}$ and  ${\bf H}_{p}(t)$  are $3\times 3$ matrices with non-zero elements  $H_{0_{ii}}=\varepsilon_{i}$, $H_{0_{23}}=H_{0_{32}}=\beta$, and $H_{p_{12}}(t)=M(t)$ is the pulse-induced light-donor coupling.
The time-dependent electromagnetic field is included in $M(t)$ that couples the highest occupied molecular orbital (HOMO) to the lowest unoccupied molecular orbital (LUMO) of the donor. Function $M(t)$  is  defined as $M(t)=\theta(t) A(t) \cos(\omega t)$ with $\theta(t)$ the Heaviside function and $A(t)=A_{0} \exp(-(t-t_{c})^{2}/2\tau^{2})$, where $A_{0}$ is the maximum amplitude and $g(t)=\exp(-(t-t_{c})^{2}/2\tau^{2})$ the Gaussian envelope of the pulse. The full width at half maximum is given by $FWHM=1.66\tau$ and $\omega$ the central frequency of the pulse.
In the present study, we fix $\varepsilon_1=-0.7$~eV, $\varepsilon_{2,3}=+0.7$~eV, $\hbar\omega=\varepsilon_{2}-\varepsilon_{1}=1.4$~eV,  and $\tau=5.0$~fs.

%%%%%%%%%%%%%%%%%%%%%%%%%%%%%%%%%%%%%%%%%
%%%%%%%%%%%%%%%%%%%%%%%%%%%%%%%%%%%%%%%%%
\section{Results and discussion}
%%%%%%%%%%%%%%%%%%%%%%%%%%%%%%%%%%%%%%%%%
%%%%%%%%%%%%%%%%%%%%%%%%%%%%%%%%%%%%%%%%%

The target is the time-resolved response of the D-A complex sandwiched between two metallic electrodes, when the donor is excited by a femtosecond Gaussian-shape laser pulse. 
The time-dependent electromagnetic field induces carrier absorption and emission from the LUMO and HOMO inside the donor. The excited electrons and holes are transferred to electrodes due to the difference of occupation between molecule and electrodes.
This results in a photocurrent through the molecular junction as a unique consequence of electron excitation by ultrafast radiation. 

\begin{figure}
   \begin{center}
	\includegraphics[width=8.25 cm]{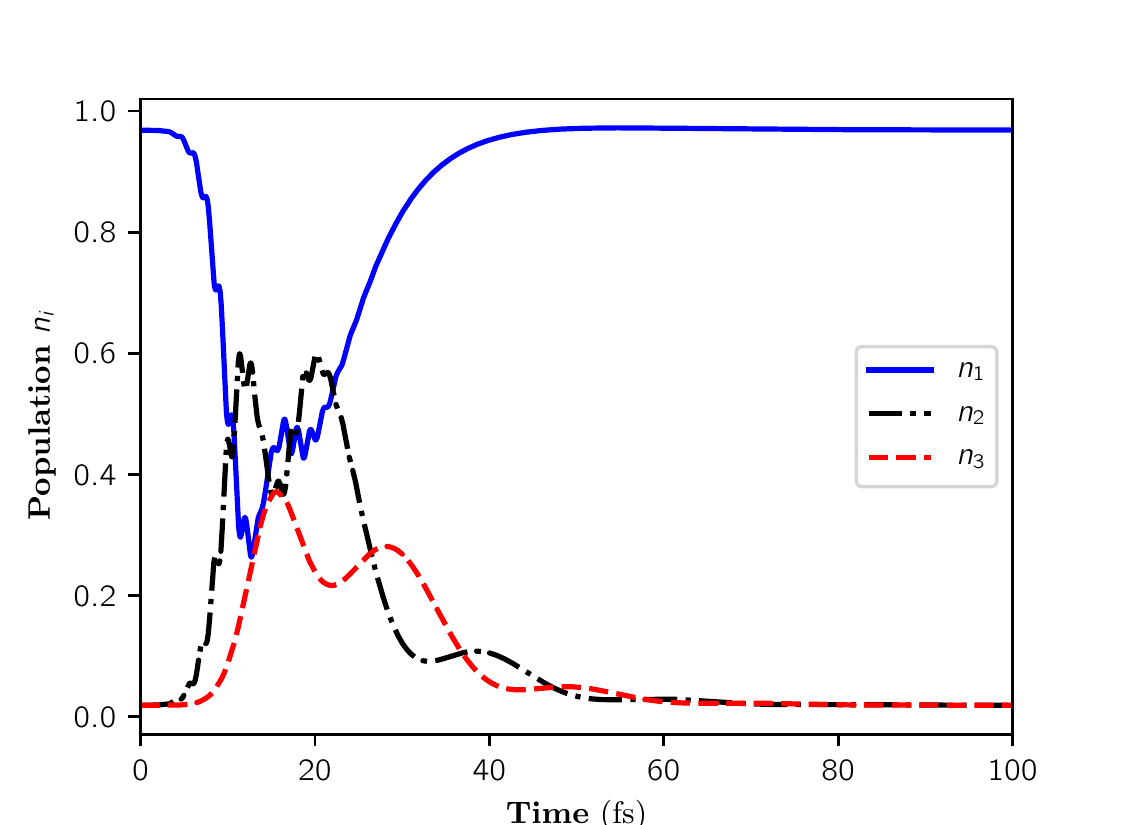}
   \end{center}
   \caption { \label{popu} Populations $n_{1}$ of donor level 1,  $n_{2}$ of donor level 2,  $n_{3}$ of acceptor level 3. Simulation parameters are $\beta=0.1$~eV, $\Gamma=0.05$~eV, and $A_{0}=0.5 eV$.}
\end{figure}

%%%% %%%%%%%%  POPULATIONS  %%%%%%%%  %%%%%%%% 
\textbf{{Population dynamics.}} 
We first analyse the dynamics of carriers inside the molecular complex.
The intramolecular orbital populations are numerically computed by integrating over energy the time-resolved spectral lesser Green's function associated to each molecular levels:
\begin{equation}
n_{i \in \{1,2,3\} }(t)=\operatorname{Im}\frac{1}{\pi}\int G_{ii}^{<}(E,t)dE  \, .
\end{equation}
Populations of Figure~\ref{popu} point out carrier pathway just after the perturbation is initiated. 
At the beginning, the pulse induces a \emph{HOMO-LUMO} transition at the donor (levels 1 to 2 of Figure~\ref{systa}), followed by intermolecular tunneling oscillations between the donor and the acceptor \emph{LUMOs} (level 2 and 3 of Figure~\ref{systa}). Meanwhile in the donor, there are noticeable interferences of propagating modes. Therefore, the pulse field induces population oscillations with two characteristic frequencies.
The first oscillation is only patterned during the light-donor interaction in populations $n_{1}$ and $n_{2}$. Corresponding frequency is about $\sim 2\omega$ ($\omega$ being the mean pulse frequency). This ``on-pulse" oscillation is the result of absorption and wave functions interferences inside the donor.  
The second oscillation occurs at the end of the pulse train: both populations $n_{2}$ and  $n_{3}$ show damped oscillations of  relaxation on Figure~\ref{popu}. These oscillations, called here ``off-pulse" oscillations for simplicity, essentially involve the two donor and acceptor LUMOs. Its frequency is controlled by both the pulse amplitude $A_{0}$ and the intramolecular donor-acceptor coupling $\beta$.
%%%%%
However, their amplitude are reduced by a high $\beta$ value, contrary to the case of the on-pulse oscillation. The amplitude of these off-pulse oscillations is strongly damped due to the relaxation of the system to equilibrium induced by the molecule-lead coupling $\Gamma$ (see again Figure~\ref{systa}).
It comes out that the simultaneous observation of on- and off-pulse oscillations could be critical due to the double role of $\beta$, but  it will depend on the choice of the bridge in building the donor-acceptor junction for carriers dynamics in molecular devices.
%%%%%
%
%\begin{figure}
%   \begin{center}
%      \begin{tabular}{c} 
%	\includegraphics[width=8.25 cm]{popudonor} \\
%      	\includegraphics[width=8.25 cm]{Field}
%     \end{tabular}
%   \end{center}
%   \caption 
%   { \label{popu} Populations (upper panel): $n_{1}$ of donor level 1,  $n_{2}$ of donor level 2,  $n_{3}$ of acceptor level 3. Simulation parameters are $\beta=0.1$~eV, $\Gamma=0.05$~eV, and $A_{0}=0.5 eV$. Pulse-induced light-donor coupling as a function of time $M(t)$ (lower panel). }
%   \end{figure} 
%

\begin{figure}
   \begin{center}
      \begin{tabular}{c} 
      \includegraphics[width=6.5 cm]{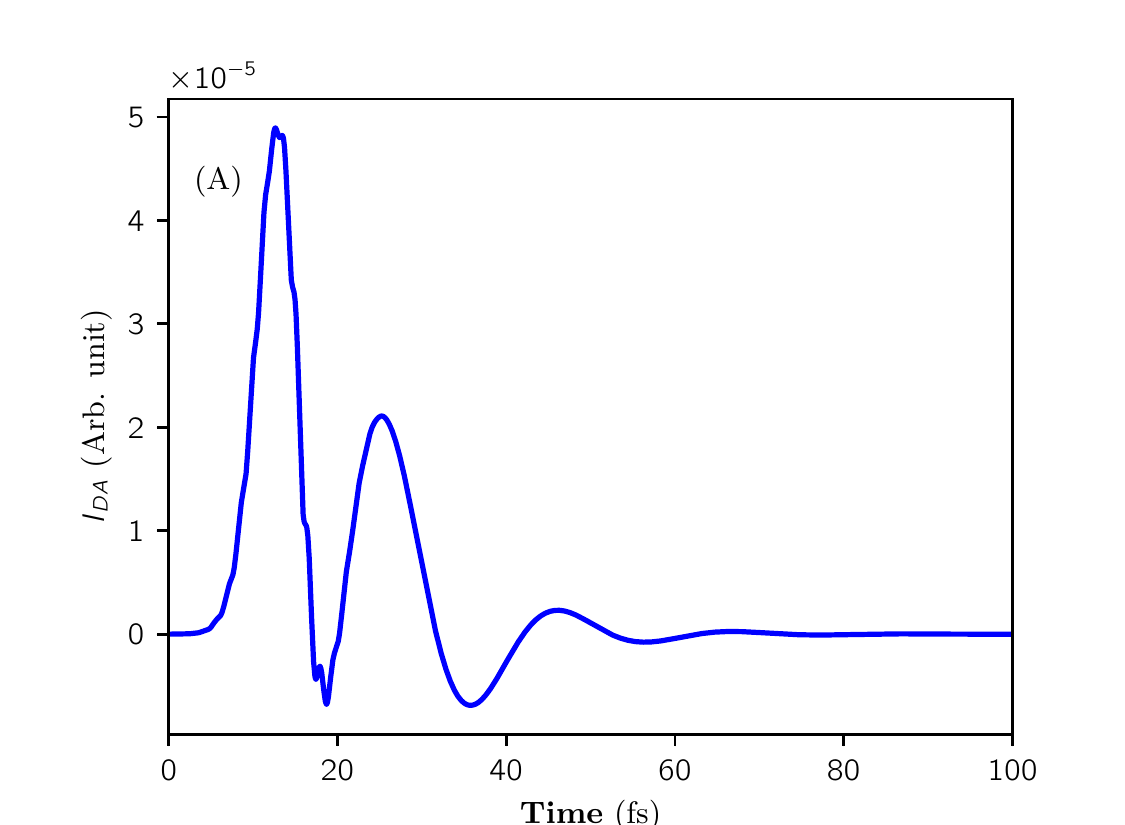} \\
      \includegraphics[width=6.5 cm]{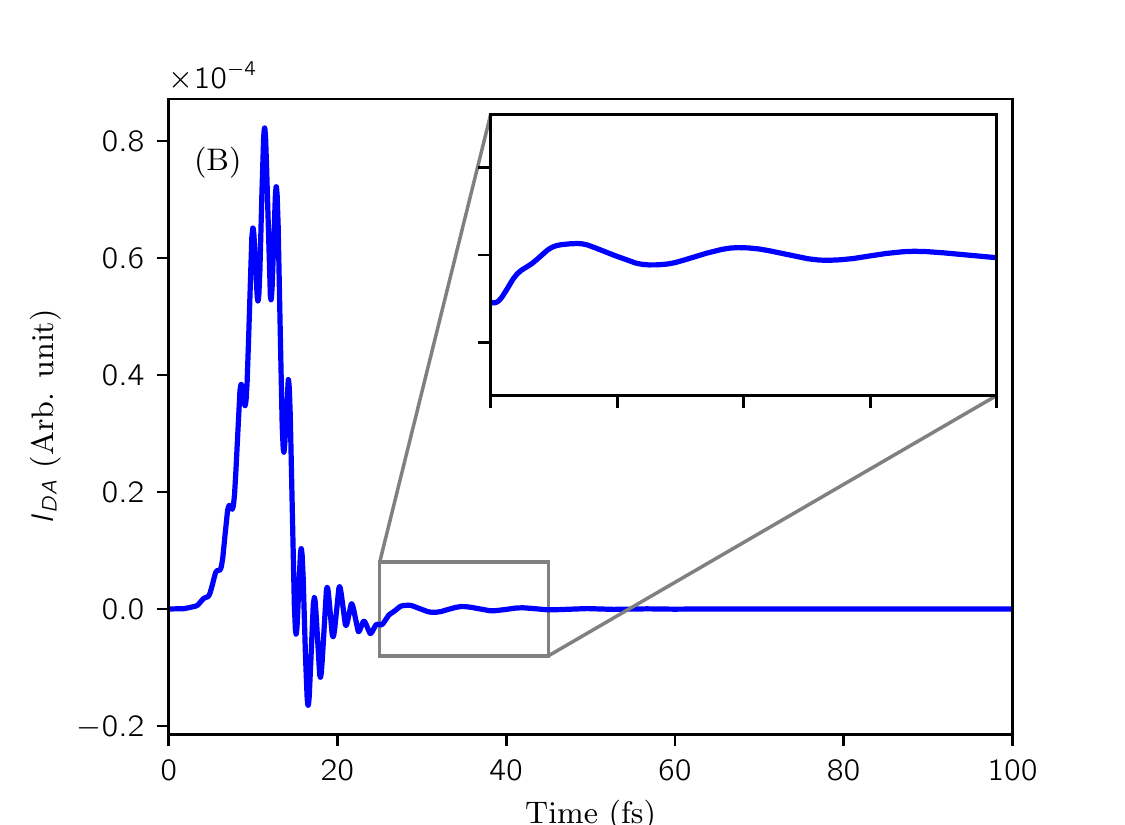} \\
      \includegraphics[width=6.5 cm]{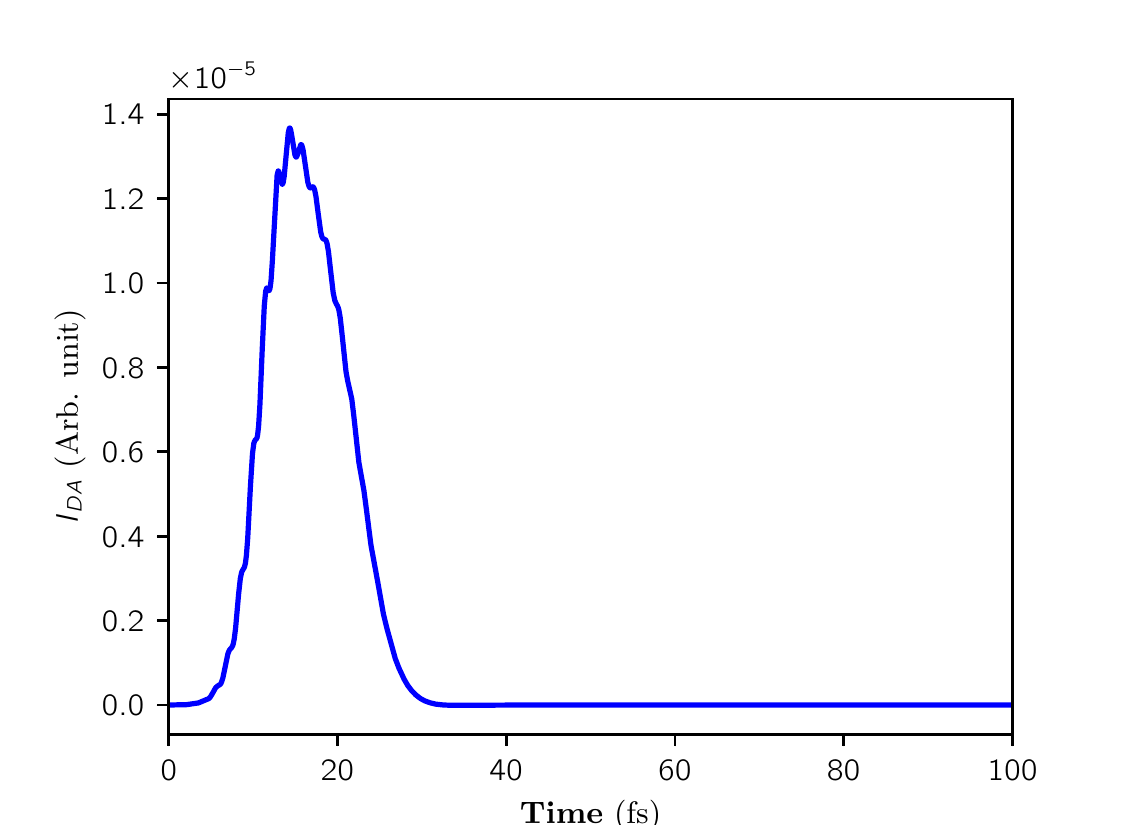} \\
      \includegraphics[width=6.5 cm]{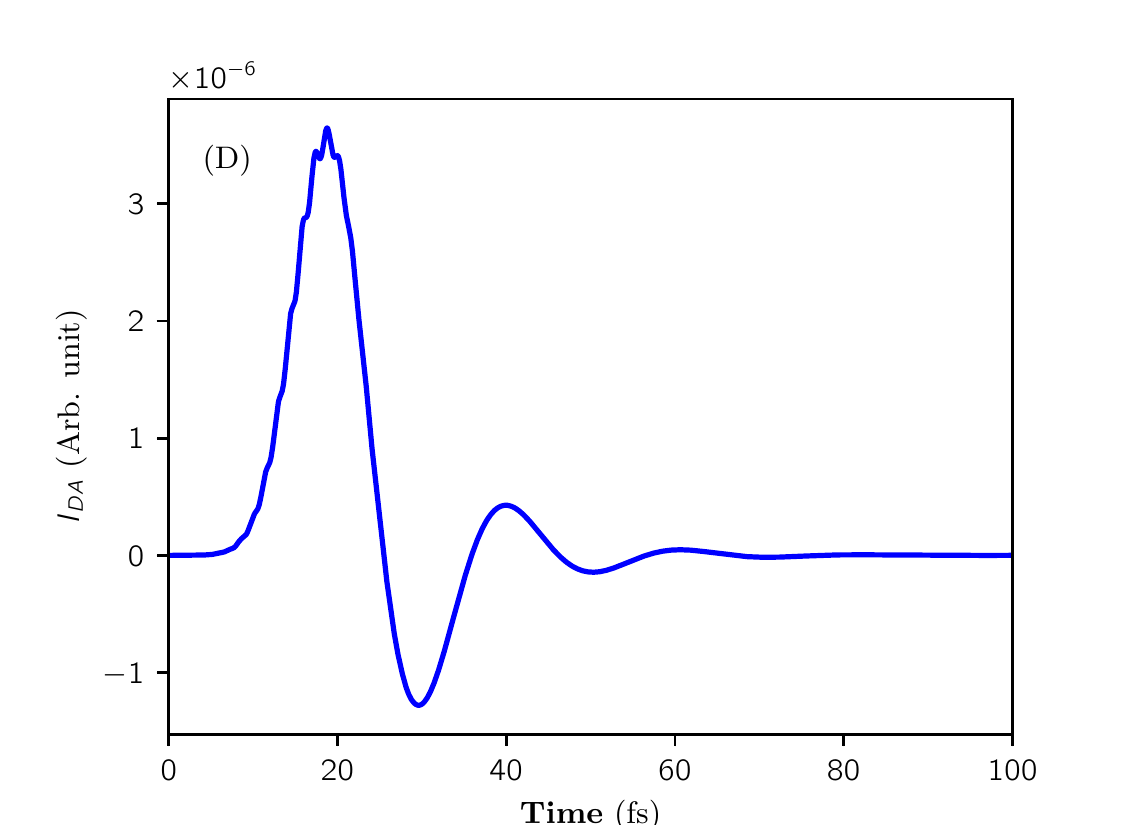}
     \end{tabular}
   \end{center}
   \caption 
   { \label{fig:currents} Time-resolved photocurrent,  (A) with the same parameters as in Figure~\ref{popu}, (B) except that $\beta=0.3$~eV instead of $0.1$~eV, (C) except that $\Gamma=0.2$~eV instead of $0.05$~eV, and (D) except that $A_0=0.05$~eV instead of $0.5$~eV. }
   \end{figure}

%%%%%%%% %%%%%%%%   CURRENT  %%%%%%%% %%%%%%%% 
\textbf{{Photocurrent generation.}}
The dynamics of populations results in a transient photocurrent within the relaxation time set by the donor-acceptor and molecule-lead couplings. 
The time-resolved photocurrent flowing from donor to acceptor $I_{DA}(t)$ reflects indeed this dynamics of populations and their characteristics, as depicted in Figure~\ref{fig:currents}(A).
Both on-pulse and off-pulse population oscillations are remarkably visible on the photocurrent variation.
%%%%%  Beta -> on pulse
In cases where on-pulse oscillations have been obtained,~\cite{croy_propagation_2009,gaury_numerical_2014} their low amplitude brought to the conclusion that they could not be  experimentally measurable due to capacitives effects.~\cite{croy_propagation_2009} However, beyond the fact that its frequency is proportional to that of the field, its relative amplitude depends on the donor-acceptor coupling strength $\beta$, as illustrated in Figure~\ref{fig:currents}(B) where $\beta$ is three times the value used in Figure~\ref{fig:currents}(A).
We thus infer that it might be possible to detect this frequency in the case of strongly connected donor-acceptor, with the right choice of bridge.
%%%%%  Beta -> off pulse
In terms of charge transfert, the off-pulse oscillation owns a characteristic frequency of forth and back tunneling between levels $\vert 2\rangle$-$\vert 3\rangle$. This frequency is related to the intramolecular D-A coupling, as can been seen in Figure~\ref{fig:currents}(B).
However,  it also can be seen that the amplitude of these intramolecular oscillations is reduced by a high $\beta$ value, contrary to the case of the on-pulse oscillation. 
It comes out that the simultaneous observation of on-pulse and off-pulse oscillations could be critical due to the double role of $\beta$, but  it will depend on the choice of the bridge in building the donor-acceptor junction for carriers dynamics in molecular devices.
%%%%%  Gamma
 The off-pulse oscillation amplitude is also strongly damped due to the relaxation of the system to equilibrium induced by the molecule-lead coupling $\Gamma$ (see again Figure~\ref{systa}).
In Figure~\ref{fig:currents}(C),  $\Gamma$ is increased compared to Figure~\ref{fig:currents}(A): one can no more observe the intramolecular oscillation neither in population (not shown) nor in photocurrent. The off-pulse oscillations are progressively damped as $\Gamma$ increases, and finally disappear at strong $\Gamma$.
The intramolecular dynamics only survives when carriers are long lived in the system, which means for a weak coupling to leads or whenever the characteristic time for the proper dynamics of the molecule is smaller than the relaxation time of the open system, $\tau_{r}=\hbar/\Gamma$. 
%%%%% A0
Finally, the light-donor coupling amplitude $A_0$ has been divided by a factor ten in Figure~\ref{fig:currents}(D). Compared to Figure~\ref{fig:currents}(A), the photocurrent amplitude is also reduced by about the same factor. Moreover, the negative oscillation inside the pulse duration is amplified, which affects the directionality of current. We quantify this directionality by defining the following ratio:
\begin{eqnarray}
r&=&\frac{\int I(t)dt}{\int_{I(t)>0} I(t)dt-\int_{I(t)<0} I(t)dt}\, ,
\end{eqnarray}
whose sign indicates the current directionality: $r=1$ if $\int_{I(t)<0} I(t)dt=0$, $r=-1$ if $\int_{I(t)>0} I(t)dt=0$, and $r=0$ if $\int_{I(t)>0} I(t)dt=\int_{I(t)<0} I(t)dt$. Comparing the four device configurations of Figure~\ref{fig:currents}(A,B,C,D), we obtain $r=0.76$ in Figure~\ref{fig:currents}(A), $0.82$ in Figure~\ref{fig:currents}(B),  $r=1.00$ in Figure~\ref{fig:currents}(C), and finally $0.58$ in Figure~\ref{fig:currents}(D). Such a control over the inversion and suppression of current was already pointed out in classical single level tunneling structure.~\cite{kuperman_field-induced_2017} We here confirm the crucial role of the field amplitude inside the light-donor coupling $M(t)$, as well as the D-A intermolecular coupling. For strong coupling to reservoir, the inversion of $I(t)$ is suppressed. Electrons do not spend enough time inside the molecule to oscillate between levels $\vert 2\rangle$-$\vert 3\rangle$ and experience strong induced emission.
This discussion on carriers dynamics and photocurrent suggests that we can generate, shape and control signals of different frequency through the design of molecular junctions.

%
%\begin{figure}
%\includegraphics[scale=0.12]{testfig1}
%\caption{Time-resolved photocurrent (upper panel), with the same parameters as in Figure~\ref{popu}. Pulse-induced light-donor coupling as a function of time $M(t)$ (lower panel).}
%\end{figure}
%
%\begin{figure}
%\includegraphics[scale=0.17]{i2.eps}
%\includegraphics[scale=0.12]{c2}
%\includegraphics[width=8.25 cm]{c2}
%\caption{\label{i2} Time-resolved photocurrent (upper panel), with the same parameters as in Figure~\ref{popu} except that $\Gamma=0.2$~eV instead of $\Gamma=0.05$~eV.  Pulse-induced light-donor coupling as a function of time $M(t)$ (lower panel).}
%\end{figure}

\begin{figure}
\includegraphics[width=8.25 cm]{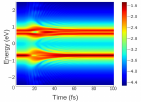}
\caption{\label{doda} Time-resolved local density of states at the donor (in logarithm scale), with the same parameters as in Figure~\ref{popu}. }
\end{figure}

%%%%%%%% %%%%%%%%  TRLDOS  %%%%%%%% %%%%%%%% 
\textbf{{Dynamical photoelectronic structure.}} 
In order to deepen the analysis, we have calculated a time-resolved local density of states (TRLDOS) at site $i$:
\begin{equation}
TRLDOS_{i}(E,t)=\operatorname{Im}\sum_{\alpha}\Big[\Psi_{\alpha E}(i,t) \Psi_{\alpha E}^{\dagger}(i,t)\Big]\, .
\end{equation}
Such a time- and energy-resolved quantity provides insight into the dynamical photoelectronic structure produced by the time-dependent electromagnetic field, as shown Figure~\ref{doda}. 

\begin{figure}
\includegraphics[width=7.0 cm]{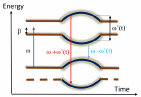}
\caption{\label{rabi} Sketch of the dynamical Rabi sidebands.}
\end{figure}

Firstly we observe a time-dependent energy level shift, known as dynamical Rabi shift.
Indeed, in a quantum system driven by an external field of time-dependent amplitude, the near-degenerated dressed states undergo time-dependent splitting $\hbar \omega '(t)=\hbar \sqrt{\beta^{2}+\omega_{r}^{2}(t)}$ where $\omega_{r}(t)$ is the instantaneous Rabi frequency given by $\omega_{r}(t)=\mu A(t)/\hbar$. This dynamical Rabi shift  generates redshifted and blueshifted sidebands at instantaneous frequencies $\omega-\omega'(t)$ and $\omega+\omega'(t)$ which are schematically drawn on Figure~\ref{rabi}.
Actually in Figure~ \ref{doda}, the $TRLDOS_D$ shows the three stationary molecular hybridized orbitals before the pulse is set on, which takes about $t=10$~fs. As soon as the pulse is set on, the energy levels are no more constant over time but dynamically shifted with respect to their stationary values. The shift results in dynamical Rabi sidebands: the three initial stationary states dynamically shifted, plus a lower-energy level appearing after $t=10$~fs, as illustrated in Figure~\ref{rabi}. This additional dynamical level is generated due to the fact that the pulse replicates the two upper energy levels inside the lower-energy part of the donor spectral function which then also splits into two Rabi sidebands~\cite{compton_dynamic_2011}.
Along the time axis,  oscillation generation are also visible at almost all energies.
After the pulse, for $t=40$~fs to $t=100$~fs, the system relaxes back to its stationary configuration with the three atomic orbitals surrounded by  interference patterns which are damped due to the molecule-lead coupling.%

\begin{figure}
\includegraphics[width=8.25 cm]{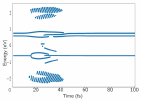}
\caption{ \label{dosu} Traces of local maxima of time-resolved local density of states extracted from Figure~\ref{doda}.}
\end{figure}

Secondly we observe lower-amplitude satellite maxima from a closer look around $\pm 2$~eV of Figure~\ref{doda}, that we identify as Floquet-like states. In order to examine these features of TRLDOS appearing in the system, we have numerically extracted and plotted the energy and time coordinates of TRLDOS local maxima, as represented Figure~\ref{dosu} (this figure also remarkably shows the dynamical Rabi shit).
We have noticed that these satellite states appear at energies about $E_i\pm\hbar \omega$, and we have checked that states follow this trend when we change $\omega$. We thus infer that these satellite states are Floquet-like states. In fact, it have been shown for a scalar periodic Hamiltonian, like in the case of applied ac-bias, that the system splits into multiple states with quasi-energies $E\pm\hbar \omega$. These Floquet states form multiple transport channels.~\cite{kuperman_field-induced_2017} In the case studied here, the electromagnetic field is not monochromatic, due to its Gaussian shape, and hence, the Hamiltonian is not periodic, but we still observe the presence of these states with almost the same quasi-energies. These Floquet-like states open new transport channels to be taken into account for transport processes, that could also control the directionality of current in molecular devices.~\cite{kuperman_field-induced_2017}. 

\begin{figure*}
   \begin{center}
      \begin{tabular}{cc} 
      \includegraphics[width=8 cm]{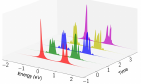} & \includegraphics[width=5 cm]{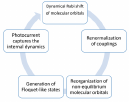}
     \end{tabular}
   \end{center}
   \caption 
   { \label{fig:ILODS} (Left) instantaneous local density of states for different times (in Arb. units), with the same parameters as in Figure~\ref{popu}. (Right) diagrammatic picture of the internal electronic structure dynamics inside the D-A molecular junction. }
   \end{figure*}

Finally, we decipher the internal dynamics of this molecular system interacting with a time-dependent electromagnetic field. 
Cuts at different times of TRLDOS are represented on the left side of Figure~\ref{fig:ILODS}.
On the time axis, we have numbered the different times: $t=0$ corresponds to the TRLDOS before excitation, $t=1,2,3$ for the TRLDOS during the excitation and $t=4$ for the TRLDOS after excitation.
These instantaneous densities of states indicate that the external field applied to the system induces a coupling of the molecular levels with the field modes, so that we not only have a rearrangement of the non-equilibrium molecular orbitals but also pulse-induced secondary maxima in the system spectral response due to the broad spectrum of the pulse, as shown on Figure~\ref{fig:ILODS}.
Due to the hybridization with states of the leads, molecular levels broaden, which allows us to describe the local molecular density of states as a sum of  level Lorentzian envelopes.
We observe that the instantaneous LDOS for $t=1,2,3$ have different full width at half maximum compared to the LDOS at $t=0$. Moreover, the energy level spacing is changed as shown up with the dynamical Rabi shift. 
The pulse tends to distort and reshape the Lorentzian envelopes, which originates from field-induced coupling renormalization in the molecular device. 
These effects of renormalization and rearrangement of molecular orbitals are suppressed for strong coupling to leads due to state delocalization. 
This last point could be a problem of experiment interpretation since the broadening could hide extra molecular orbitals, or even shifts.
The diagrammatic sum up of our interpretation of charge dynamics in this simple D-A complex device is finally illustrated in the left panel of Figure~\ref{fig:ILODS}.
   
%
% \begin{figure}[H]
%\includegraphics[scale=0.12]{trldos2}
%\caption{\label{dos3D1} Instantaneous LDOS for different times. Same parameters as in Figure~\ref{popu}.}
%\end{figure}
%
%\begin{figure}[H]
%\includegraphics[scale=0.14]{thmv2}
%\caption{\label{GFTI} }
%\end{figure}

%%%%%%%% %%%%%%%%  Discussion  %%%%%%%% %%%%%%%% 
\textbf{{Discussion.}} 
Applying an ultrashort laser pulse induces a dynamical energy level shift and generation of Rabi sidebands that induces Floquet-like states, both shown on Figure~\ref{dosu}. These Floquet-like states could be the anomalous  edge states observed in weakly driven lattice systems. The photocurrent is sensitive to the quasi-energy spectrum of the transient Floquet states and the Rabi sidebands. One could capture, generate, shape and control high frequencies from the intramolecular dynamics or the external electromagnetic field. 
The signature of the Rabi sidebands could be observed as blueshifted and redshifted frequencies in the absorption spectrum of such devices. Moreover, these sidebands also manifest variations of tunneling parameters, which provide the coherent control of the field-induced photocurrent.

However, realistic simulation including electron-electron scattering and molecular vibrations might lead to stronger damping of the oscillations described above. Electron-electron scattering nonetheless occurs on the time scale of picoseconds while
the electronic oscillations have oscillation periods of tens of femtoseconds or less. Hence, if the
ultimate goal of molecular electronics is to achieve switching times on the electronic time
scale, the oscillations predicted here will be highly relevant.
Molecular vibrations and electrostatics will be part of future works.

%%%%%%%%%%%%%%%%%%%%%%%%%%%%%%%%%%%%%%%%%
%%%%%%%%%%%%%%%%%%%%%%%%%%%%%%%%%%%%%%%%%
\section{Conclusion}
%%%%%%%%%%%%%%%%%%%%%%%%%%%%%%%%%%%%%%%%%
%%%%%%%%%%%%%%%%%%%%%%%%%%%%%%%%%%%%%%%%%

The present study attempts to establish the time-resolved response of a D-A molecular complex to a femtosecond Gaussian-shape laser pulse within the framework of Keldysh's formalism.  It relies on the analysis of the dynamical photoelectronic structure in which we have demonstrated several transient features: oscillation generation and damping, dynamical Rabi shift of energy levels and appearance of Floquet-like states. 
This work opens indeed a discussion on the relevance of transient dynamics in the understanding of time-resolved device operations, and at the same time it opens avenues towards ultrafast device design for future technologies.

%%%%%%%%%%%%%%%%%%%%%%%%%%%%%%%%%%%%%%%%%
%%%%%%%%%%%%%%%%%%%%%%%%%%%%%%%%%%%%%%%%%
\begin{acknowledgement}
%%%%%%%%%%%%%%%%%%%%%%%%%%%%%%%%%%%%%%%%%
%%%%%%%%%%%%%%%%%%%%%%%%%%%%%%%%%%%%%%%%%
The authors acknowledge financial support from the ANR-French National  Research  Agency [NOODLES, 
contract  No. ANR-13-NANO-0009].
\end{acknowledgement}

\bibliographystyle{achemso}
%\bibliography{PAPER2017}
\bibliography{bibbelt}

\providecommand{\latin}[1]{#1}
\providecommand*\mcitethebibliography{\thebibliography}
\csname @ifundefined\endcsname{endmcitethebibliography}
  {\let\endmcitethebibliography\endthebibliography}{}
\begin{mcitethebibliography}{32}
\providecommand*\natexlab[1]{#1}
\providecommand*\mciteSetBstSublistMode[1]{}
\providecommand*\mciteSetBstMaxWidthForm[2]{}
\providecommand*\mciteBstWouldAddEndPuncttrue
  {\def\EndOfBibitem{\unskip.}}
\providecommand*\mciteBstWouldAddEndPunctfalse
  {\let\EndOfBibitem\relax}
\providecommand*\mciteSetBstMidEndSepPunct[3]{}
\providecommand*\mciteSetBstSublistLabelBeginEnd[3]{}
\providecommand*\EndOfBibitem{}
\mciteSetBstSublistMode{f}
\mciteSetBstMaxWidthForm{subitem}{(\alph{mcitesubitemcount})}
\mciteSetBstSublistLabelBeginEnd
  {\mcitemaxwidthsubitemform\space}
  {\relax}
  {\relax}

\bibitem[Mentovich \latin{et~al.}(2013)Mentovich, Rosenberg-Shraga, Kalifa,
  Gozin, Mujica, Hansen, and Richter]{mentovich_gated-controlled_2013}
Mentovich,~E.~D.; Rosenberg-Shraga,~N.; Kalifa,~I.; Gozin,~M.; Mujica,~V.;
  Hansen,~T.; Richter,~S. \emph{J. Phys. Chem. C} \textbf{2013}, \emph{117},
  8468--8474\relax
\mciteBstWouldAddEndPuncttrue
\mciteSetBstMidEndSepPunct{\mcitedefaultmidpunct}
{\mcitedefaultendpunct}{\mcitedefaultseppunct}\relax
\EndOfBibitem
\bibitem[Díez-Pérez \latin{et~al.}(2012)Díez-Pérez, Li, Guo, Madden, Huang,
  Che, Yang, Zang, and Tao]{diez-perez_ambipolar_2012}
Díez-Pérez,~I.; Li,~Z.; Guo,~S.; Madden,~C.; Huang,~H.; Che,~Y.; Yang,~X.;
  Zang,~L.; Tao,~N. \emph{ACS Nano} \textbf{2012}, \emph{6}, 7044--7052\relax
\mciteBstWouldAddEndPuncttrue
\mciteSetBstMidEndSepPunct{\mcitedefaultmidpunct}
{\mcitedefaultendpunct}{\mcitedefaultseppunct}\relax
\EndOfBibitem
\bibitem[{Xu} \latin{et~al.}(2005){Xu}, {Xiao}, Yang, Zang, and
  {Tao}]{xu_large_2005}
{Xu},; {Xiao},; Yang,~X.; Zang,~L.; {Tao}, \emph{J. Am. Chem. Soc.}
  \textbf{2005}, \emph{127}, 2386--2387\relax
\mciteBstWouldAddEndPuncttrue
\mciteSetBstMidEndSepPunct{\mcitedefaultmidpunct}
{\mcitedefaultendpunct}{\mcitedefaultseppunct}\relax
\EndOfBibitem
\bibitem[Lörtscher \latin{et~al.}(2012)Lörtscher, Gotsmann, Lee, Yu, Rettner,
  and Riel]{lortscher_transport_2012}
Lörtscher,~E.; Gotsmann,~B.; Lee,~Y.; Yu,~L.; Rettner,~C.; Riel,~H. \emph{ACS
  Nano} \textbf{2012}, \emph{6}, 4931--4939\relax
\mciteBstWouldAddEndPuncttrue
\mciteSetBstMidEndSepPunct{\mcitedefaultmidpunct}
{\mcitedefaultendpunct}{\mcitedefaultseppunct}\relax
\EndOfBibitem
\bibitem[Selzer and Peskin(2013)Selzer, and Peskin]{selzer_transient_2013}
Selzer,~Y.; Peskin,~U. \emph{J. Phys. Chem. C} \textbf{2013}, \emph{117},
  22369--22376\relax
\mciteBstWouldAddEndPuncttrue
\mciteSetBstMidEndSepPunct{\mcitedefaultmidpunct}
{\mcitedefaultendpunct}{\mcitedefaultseppunct}\relax
\EndOfBibitem
\bibitem[Volkovich and Peskin(2011)Volkovich, and
  Peskin]{volkovich_transient_2011}
Volkovich,~R.; Peskin,~U. \emph{Phys. Rev. B} \textbf{2011}, \emph{83},
  033403\relax
\mciteBstWouldAddEndPuncttrue
\mciteSetBstMidEndSepPunct{\mcitedefaultmidpunct}
{\mcitedefaultendpunct}{\mcitedefaultseppunct}\relax
\EndOfBibitem
\bibitem[Platero and Aguado(2004)Platero, and
  Aguado]{platero_photon-assisted_2004}
Platero,~G.; Aguado,~R. \emph{Physics Reports} \textbf{2004}, \emph{395},
  1--157\relax
\mciteBstWouldAddEndPuncttrue
\mciteSetBstMidEndSepPunct{\mcitedefaultmidpunct}
{\mcitedefaultendpunct}{\mcitedefaultseppunct}\relax
\EndOfBibitem
\bibitem[Cocker \latin{et~al.}(2013)Cocker, Jelic, Gupta, Molesky, Burgess,
  Reyes, Titova, Tsui, Freeman, and Hegmann]{cocker_ultrafast_2013}
Cocker,~T.~L.; Jelic,~V.; Gupta,~M.; Molesky,~S.~J.; Burgess,~J. A.~J.;
  Reyes,~G. D.~L.; Titova,~L.~V.; Tsui,~Y.~Y.; Freeman,~M.~R.; Hegmann,~F.~A.
  \emph{Nat Photon} \textbf{2013}, \emph{7}, 620--625\relax
\mciteBstWouldAddEndPuncttrue
\mciteSetBstMidEndSepPunct{\mcitedefaultmidpunct}
{\mcitedefaultendpunct}{\mcitedefaultseppunct}\relax
\EndOfBibitem
\bibitem[Pivrikas \latin{et~al.}(2007)Pivrikas, Sariciftci, Juška, and
  Österbacka]{pivrikas_review_2007}
Pivrikas,~A.; Sariciftci,~N.~S.; Juška,~G.; Österbacka,~R. \emph{Prog.
  Photovolt: Res. Appl.} \textbf{2007}, \emph{15}, 677--696\relax
\mciteBstWouldAddEndPuncttrue
\mciteSetBstMidEndSepPunct{\mcitedefaultmidpunct}
{\mcitedefaultendpunct}{\mcitedefaultseppunct}\relax
\EndOfBibitem
\bibitem[Brauer \latin{et~al.}(2015)Brauer, Marchioro, Paraecattil, Oskouei,
  and Moser]{brauer_dynamics_2015}
Brauer,~J.~C.; Marchioro,~A.; Paraecattil,~A.~A.; Oskouei,~A.~A.; Moser,~J.-E.
  \emph{J. Phys. Chem. C} \textbf{2015}, \emph{119}, 26266--26274\relax
\mciteBstWouldAddEndPuncttrue
\mciteSetBstMidEndSepPunct{\mcitedefaultmidpunct}
{\mcitedefaultendpunct}{\mcitedefaultseppunct}\relax
\EndOfBibitem
\bibitem[Bakulin \latin{et~al.}(2013)Bakulin, Neutzner, Bakker, Ottaviani,
  Barakel, and Chen]{bakulin_charge_2013}
Bakulin,~A.~A.; Neutzner,~S.; Bakker,~H.~J.; Ottaviani,~L.; Barakel,~D.;
  Chen,~Z. \emph{ACS Nano} \textbf{2013}, \emph{7}, 8771--8779\relax
\mciteBstWouldAddEndPuncttrue
\mciteSetBstMidEndSepPunct{\mcitedefaultmidpunct}
{\mcitedefaultendpunct}{\mcitedefaultseppunct}\relax
\EndOfBibitem
\bibitem[Bakulin \latin{et~al.}(2016)Bakulin, Silva, and
  Vella]{bakulin_ultrafast_2016}
Bakulin,~A.~A.; Silva,~C.; Vella,~E. \emph{J. Phys. Chem. Lett.} \textbf{2016},
  \emph{7}, 250--258\relax
\mciteBstWouldAddEndPuncttrue
\mciteSetBstMidEndSepPunct{\mcitedefaultmidpunct}
{\mcitedefaultendpunct}{\mcitedefaultseppunct}\relax
\EndOfBibitem
\bibitem[Jakowetz \latin{et~al.}(2016)Jakowetz, Böhm, Zhang, Sadhanala,
  Huettner, Bakulin, Rao, and Friend]{jakowetz_what_2016}
Jakowetz,~A.~C.; Böhm,~M.~L.; Zhang,~J.; Sadhanala,~A.; Huettner,~S.;
  Bakulin,~A.~A.; Rao,~A.; Friend,~R.~H. \emph{J. Am. Chem. Soc.}
  \textbf{2016}, \emph{138}, 11672--11679\relax
\mciteBstWouldAddEndPuncttrue
\mciteSetBstMidEndSepPunct{\mcitedefaultmidpunct}
{\mcitedefaultendpunct}{\mcitedefaultseppunct}\relax
\EndOfBibitem
\bibitem[Ono and Ohno(2016)Ono, and Ohno]{ono_minimal_2016}
Ono,~S.; Ohno,~K. \emph{Phys. Rev. B} \textbf{2016}, \emph{93}, 121301\relax
\mciteBstWouldAddEndPuncttrue
\mciteSetBstMidEndSepPunct{\mcitedefaultmidpunct}
{\mcitedefaultendpunct}{\mcitedefaultseppunct}\relax
\EndOfBibitem
\bibitem[Prins \latin{et~al.}(2012)Prins, Buscema, Seldenthuis, Etaki, Buchs,
  Barkelid, Zwiller, Gao, Houtepen, Siebbeles, and van~der
  Zant]{prins_fast_2012}
Prins,~F.; Buscema,~M.; Seldenthuis,~J.~S.; Etaki,~S.; Buchs,~G.; Barkelid,~M.;
  Zwiller,~V.; Gao,~Y.; Houtepen,~A.~J.; Siebbeles,~L. D.~A.; van~der Zant,~H.
  S.~J. \emph{Nano Lett.} \textbf{2012}, \emph{12}, 5740--5743\relax
\mciteBstWouldAddEndPuncttrue
\mciteSetBstMidEndSepPunct{\mcitedefaultmidpunct}
{\mcitedefaultendpunct}{\mcitedefaultseppunct}\relax
\EndOfBibitem
\bibitem[Alsulami \latin{et~al.}(2016)Alsulami, Murali, Alsinan, Parida, Aly,
  and Mohammed]{alsulami_remarkably_2016}
Alsulami,~Q.~A.; Murali,~B.; Alsinan,~Y.; Parida,~M.~R.; Aly,~S.~M.;
  Mohammed,~O.~F. \emph{Adv. Energy Mater.} \textbf{2016}, \emph{6},
  n/a--n/a\relax
\mciteBstWouldAddEndPuncttrue
\mciteSetBstMidEndSepPunct{\mcitedefaultmidpunct}
{\mcitedefaultendpunct}{\mcitedefaultseppunct}\relax
\EndOfBibitem
\bibitem[Cavassilas \latin{et~al.}(2013)Cavassilas, Michelini, and
  Bescond]{cavassilas_modeling_2013}
Cavassilas,~N.; Michelini,~F.; Bescond,~M. \emph{Journal of Renewable and
  Sustainable Energy} \textbf{2013}, \emph{6}, 011203\relax
\mciteBstWouldAddEndPuncttrue
\mciteSetBstMidEndSepPunct{\mcitedefaultmidpunct}
{\mcitedefaultendpunct}{\mcitedefaultseppunct}\relax
\EndOfBibitem
\bibitem[Beltako \latin{et~al.}(2016)Beltako, Cavassilas, and
  Michelini]{beltako_state_2016}
Beltako,~K.; Cavassilas,~N.; Michelini,~F. \emph{Appl. Phys. Lett.}
  \textbf{2016}, \emph{109}, 073501, WOS:000383787400045\relax
\mciteBstWouldAddEndPuncttrue
\mciteSetBstMidEndSepPunct{\mcitedefaultmidpunct}
{\mcitedefaultendpunct}{\mcitedefaultseppunct}\relax
\EndOfBibitem
\bibitem[Michelini \latin{et~al.}(2017)Michelini, Crépieux, and
  Beltako]{michelini_entropy_2017}
Michelini,~F.; Crépieux,~A.; Beltako,~K. \emph{J. Phys.: Condens. Matter}
  \textbf{2017}, \emph{29}, 175301\relax
\mciteBstWouldAddEndPuncttrue
\mciteSetBstMidEndSepPunct{\mcitedefaultmidpunct}
{\mcitedefaultendpunct}{\mcitedefaultseppunct}\relax
\EndOfBibitem
\bibitem[Nemati~Aram \latin{et~al.}(2016)Nemati~Aram, Anghel-Vasilescu, Asgari,
  Ernzerhof, and Mayou]{nemati_aram_modeling_2016}
Nemati~Aram,~T.; Anghel-Vasilescu,~P.; Asgari,~A.; Ernzerhof,~M.; Mayou,~D.
  \emph{The Journal of Chemical Physics} \textbf{2016}, \emph{145},
  124116\relax
\mciteBstWouldAddEndPuncttrue
\mciteSetBstMidEndSepPunct{\mcitedefaultmidpunct}
{\mcitedefaultendpunct}{\mcitedefaultseppunct}\relax
\EndOfBibitem
\bibitem[Nemati~Aram \latin{et~al.}(2017)Nemati~Aram, Ernzerhof, Asgari, and
  Mayou]{nemati_aram_impact_2017}
Nemati~Aram,~T.; Ernzerhof,~M.; Asgari,~A.; Mayou,~D. \emph{The Journal of
  Chemical Physics} \textbf{2017}, \emph{146}, 034103\relax
\mciteBstWouldAddEndPuncttrue
\mciteSetBstMidEndSepPunct{\mcitedefaultmidpunct}
{\mcitedefaultendpunct}{\mcitedefaultseppunct}\relax
\EndOfBibitem
\bibitem[Xiang \latin{et~al.}(2016)Xiang, Wang, Jia, Lee, and
  Guo]{xiang_molecular-scale_2016}
Xiang,~D.; Wang,~X.; Jia,~C.; Lee,~T.; Guo,~X. \emph{Chem. Rev.} \textbf{2016},
  \emph{116}, 4318--4440\relax
\mciteBstWouldAddEndPuncttrue
\mciteSetBstMidEndSepPunct{\mcitedefaultmidpunct}
{\mcitedefaultendpunct}{\mcitedefaultseppunct}\relax
\EndOfBibitem
\bibitem[Evers and Venkataraman(2017)Evers, and
  Venkataraman]{evers_preface:_2017}
Evers,~F.; Venkataraman,~L. \emph{The Journal of Chemical Physics}
  \textbf{2017}, \emph{146}, 092101\relax
\mciteBstWouldAddEndPuncttrue
\mciteSetBstMidEndSepPunct{\mcitedefaultmidpunct}
{\mcitedefaultendpunct}{\mcitedefaultseppunct}\relax
\EndOfBibitem
\bibitem[Kuperman and Peskin(2017)Kuperman, and
  Peskin]{kuperman_field-induced_2017}
Kuperman,~M.; Peskin,~U. \emph{The Journal of Chemical Physics} \textbf{2017},
  \emph{146}, 092314\relax
\mciteBstWouldAddEndPuncttrue
\mciteSetBstMidEndSepPunct{\mcitedefaultmidpunct}
{\mcitedefaultendpunct}{\mcitedefaultseppunct}\relax
\EndOfBibitem
\bibitem[Hertel and Radloff(2006)Hertel, and Radloff]{hertel_ultrafast_2006}
Hertel,~I.~V.; Radloff,~W. \emph{Rep. Prog. Phys.} \textbf{2006}, \emph{69},
  1897\relax
\mciteBstWouldAddEndPuncttrue
\mciteSetBstMidEndSepPunct{\mcitedefaultmidpunct}
{\mcitedefaultendpunct}{\mcitedefaultseppunct}\relax
\EndOfBibitem
\bibitem[Compton \latin{et~al.}(2011)Compton, Filin, Romanov, and
  Levis]{compton_dynamic_2011}
Compton,~R.; Filin,~A.; Romanov,~D.~A.; Levis,~R.~J. \emph{Phys. Rev. A}
  \textbf{2011}, \emph{83}, 053423\relax
\mciteBstWouldAddEndPuncttrue
\mciteSetBstMidEndSepPunct{\mcitedefaultmidpunct}
{\mcitedefaultendpunct}{\mcitedefaultseppunct}\relax
\EndOfBibitem
\bibitem[Cochrane \latin{et~al.}(2015)Cochrane, Schiffrin, Roussy, Capsoni, and
  Burke]{cochrane_pronounced_2015}
Cochrane,~K.~A.; Schiffrin,~A.; Roussy,~T.~S.; Capsoni,~M.; Burke,~S.~A.
  \emph{Nature Communications} \textbf{2015}, \emph{6}, 8312\relax
\mciteBstWouldAddEndPuncttrue
\mciteSetBstMidEndSepPunct{\mcitedefaultmidpunct}
{\mcitedefaultendpunct}{\mcitedefaultseppunct}\relax
\EndOfBibitem
\bibitem[Lake and Datta(1992)Lake, and Datta]{lake_nonequilibrium_1992}
Lake,~R.; Datta,~S. \emph{Phys. Rev. B} \textbf{1992}, \emph{45},
  6670--6685\relax
\mciteBstWouldAddEndPuncttrue
\mciteSetBstMidEndSepPunct{\mcitedefaultmidpunct}
{\mcitedefaultendpunct}{\mcitedefaultseppunct}\relax
\EndOfBibitem
\bibitem[Gaury \latin{et~al.}(2014)Gaury, Weston, Santin, Houzet, Groth, and
  Waintal]{gaury_numerical_2014}
Gaury,~B.; Weston,~J.; Santin,~M.; Houzet,~M.; Groth,~C.; Waintal,~X.
  \emph{Physics Reports} \textbf{2014}, \emph{534}, 1--37\relax
\mciteBstWouldAddEndPuncttrue
\mciteSetBstMidEndSepPunct{\mcitedefaultmidpunct}
{\mcitedefaultendpunct}{\mcitedefaultseppunct}\relax
\EndOfBibitem
\bibitem[Settnes \latin{et~al.}(2015)Settnes, Power, Lin, Petersen, and
  Jauho]{settnes_patched_2015}
Settnes,~M.; Power,~S.~R.; Lin,~J.; Petersen,~D.~H.; Jauho,~A.-P. \emph{Phys.
  Rev. B} \textbf{2015}, \emph{91}, 125408\relax
\mciteBstWouldAddEndPuncttrue
\mciteSetBstMidEndSepPunct{\mcitedefaultmidpunct}
{\mcitedefaultendpunct}{\mcitedefaultseppunct}\relax
\EndOfBibitem
\bibitem[Croy and Saalmann(2009)Croy, and Saalmann]{croy_propagation_2009}
Croy,~A.; Saalmann,~U. \emph{Phys. Rev. B} \textbf{2009}, \emph{80},
  245311\relax
\mciteBstWouldAddEndPuncttrue
\mciteSetBstMidEndSepPunct{\mcitedefaultmidpunct}
{\mcitedefaultendpunct}{\mcitedefaultseppunct}\relax
\EndOfBibitem
\end{mcitethebibliography}

\end{document}